\documentclass[twoside,twocolumn,9pt]{article}
\usepackage{extsizes}
\usepackage[super,sort&compress,comma]{natbib} 
\usepackage[version=3]{mhchem}
\usepackage[left=1.5cm, right=1.5cm, top=1.785cm, bottom=2.0cm]{geometry}
\usepackage{balance}
\usepackage{mathptmx}
\usepackage{sectsty}
\usepackage{graphicx} 
\usepackage{lastpage}
\usepackage[format=plain,justification=justified,singlelinecheck=false,font={stretch=1.125,small,sf},labelfont=bf,labelsep=space]{caption}
\usepackage{float}
\usepackage{fancyhdr}
\usepackage{fnpos}
\usepackage[english]{babel}
\addto{\captionsenglish}{%
  
}
\usepackage{array}
\usepackage{droidsans}
\usepackage{charter}
\usepackage[T1]{fontenc}
\usepackage[usenames,dvipsnames]{xcolor}
\usepackage{setspace}
\usepackage[compact]{titlesec}
\usepackage{hyperref}

\usepackage{amsmath, amssymb}
\usepackage{mathtools}

\usepackage{epstopdf}

\definecolor{cream}{RGB}{222,217,201}

\begin{document}

\pagestyle{fancy}
\thispagestyle{plain}
\fancypagestyle{plain}{
\renewcommand{\headrulewidth}{0pt}
}

\makeFNbottom
\makeatletter
\renewcommand\LARGE{\@setfontsize\LARGE{15pt}{17}}
\renewcommand\Large{\@setfontsize\Large{12pt}{14}}
\renewcommand\large{\@setfontsize\large{10pt}{12}}
\renewcommand\footnotesize{\@setfontsize\footnotesize{7pt}{10}}
\makeatother

\renewcommand{\thefootnote}{\fnsymbol{footnote}}
\renewcommand\footnoterule{\vspace*{1pt}%
\color{cream}\hrule width 3.5in height 0.4pt \color{black}\vspace*{5pt}} 
\setcounter{secnumdepth}{5}

\makeatletter 
\renewcommand\@biblabel[1]{#1}            
\renewcommand\@makefntext[1]%
{\noindent\makebox[0pt][r]{\@thefnmark\,}#1}
\makeatother 
\renewcommand{\figurename}{\small{Fig.}~}
\sectionfont{\sffamily\Large}
\subsectionfont{\normalsize}
\subsubsectionfont{\bf}
\setstretch{1.125} 
\setlength{\skip\footins}{0.8cm}
\setlength{\footnotesep}{0.25cm}
\setlength{\jot}{10pt}
\titlespacing*{\section}{0pt}{4pt}{4pt}
\titlespacing*{\subsection}{0pt}{15pt}{1pt}

\fancyfoot{}
\fancyfoot[LO,RE]{\vspace{-7.1pt}\includegraphics[height=9pt]{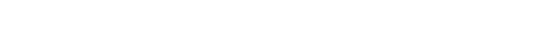}}
\fancyfoot[CO]{\vspace{-7.1pt}\hspace{13.2cm}\includegraphics{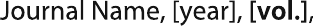}}
\fancyfoot[CE]{\vspace{-7.2pt}\hspace{-14.2cm}\includegraphics{head_foot/RF}}
\fancyfoot[RO]{\footnotesize{\sffamily{1--\pageref{LastPage} ~\textbar  \hspace{2pt}\thepage}}}
\fancyfoot[LE]{\footnotesize{\sffamily{\thepage~\textbar\hspace{3.45cm} 1--\pageref{LastPage}}}}
\fancyhead{}
\renewcommand{\headrulewidth}{0pt} 
\renewcommand{\footrulewidth}{0pt}
\setlength{\arrayrulewidth}{1pt}
\setlength{\columnsep}{6.5mm}
\setlength\bibsep{1pt}

\makeatletter 
\newlength{\figrulesep} 
\setlength{\figrulesep}{0.5\textfloatsep} 

\newcommand{\topfigrule}{\vspace*{-1pt}%
\noindent{\color{cream}\rule[-\figrulesep]{\columnwidth}{1.5pt}} }

\newcommand{\botfigrule}{\vspace*{-2pt}%
\noindent{\color{cream}\rule[\figrulesep]{\columnwidth}{1.5pt}} }

\newcommand{\dblfigrule}{\vspace*{-1pt}%
\noindent{\color{cream}\rule[-\figrulesep]{\textwidth}{1.5pt}} }

\makeatother

\twocolumn[
  \begin{@twocolumnfalse}
{\includegraphics[height=30pt]{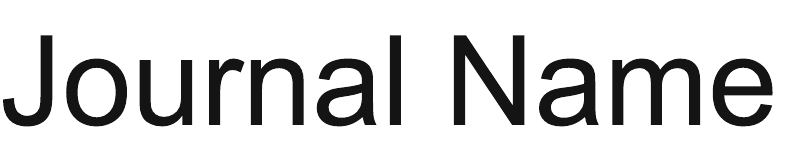}\hfill\raisebox{0pt}[0pt][0pt]{\includegraphics[height=55pt]{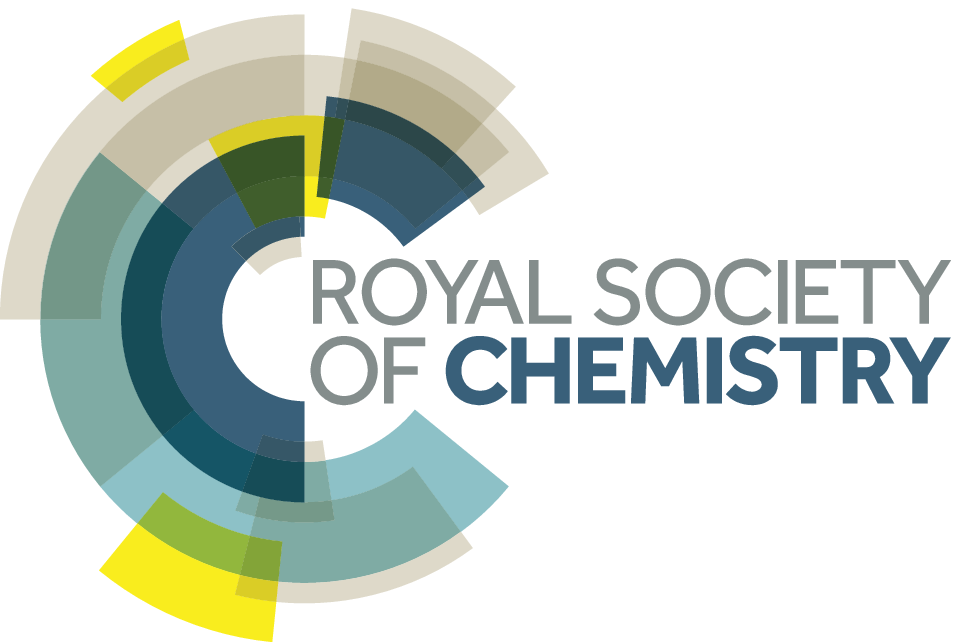}}\\[1ex]
\includegraphics[width=18.5cm]{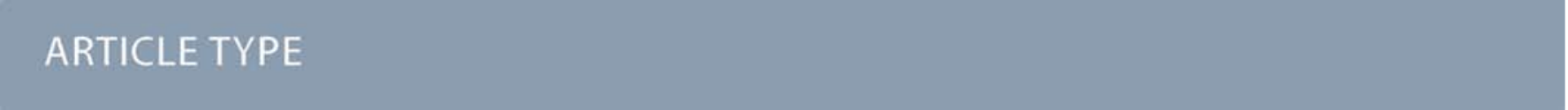}}\par
\vspace{1em}
\sffamily
\begin{tabular}{m{4.5cm} p{13.5cm} }

\includegraphics{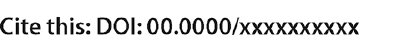} & \noindent\LARGE{\textbf{Experimental evidence of detailed balance in granular systems}} \\
\vspace{0.3cm} & \vspace{0.3cm} \\

 & \noindent\large{Xulai Sun,\textit{$^{a}$} Yinqiao Wang,\textit{$^{a}$} Yujie Wang,\textit{$^{a}$} Raphael Blumenfeld,$^{\ast}$\textit{$^{b}$} and Jie Zhang$^{\ast}$\textit{$^{a,c}$}} \\


\includegraphics{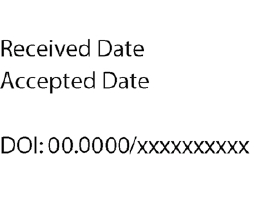} & \noindent\normalsize{The principle of detailed balance (DB) states that every kinetic transition in a system with many micro-states, $\mu$, is balanced, on average, with the opposite transition, $\mu_i\leftrightharpoons\mu_j$. Since its introduction by Boltzmann, this principle has been used by luminaries, such as Einstein, Eddington, Kramers, Pauli, Ehrenfest, Dirac, Onsager, and many others to derive significant results that underpin much of our scientific understanding. The current belief is that DB is satisfied only in equilibrium systems, while non-equilibrium steady states can only be balanced by cycles, such as $A\to B\to C\to A$. We show here experimentally that DB can exist and is commonly and robustly satisfied in a family of quasi-statically cyclically sheared granular systems.  We further study the approach to DB as a function of system size and time. Given the significant impact that this principle has had on equilibrium systems, we believe that this discovery paves the way for better models of the dynamics of non-equilibrium systems.} \\

\end{tabular}

 \end{@twocolumnfalse} \vspace{0.6cm}

  ]

\renewcommand*\rmdefault{bch}\normalfont\upshape
\rmfamily
\section*{}
\vspace{-1cm}


\footnotetext{\textit{$^{a}$~School of Physics and Astronomy, Shanghai Jiao Tong University, 800 Dong Chuan Road, Shanghai, China.}}
\footnotetext{\textit{$^{b}$~Gonville \& Caius College, University of Cambridge, Cambridge, UK. E-mail: rbb11@cam.ac.uk}}
\footnotetext{\textit{$^{a}$~Institute of Natural Sciences, Shanghai Jiao Tong University, 800 Dong Chuan Road, Shanghai, China. E-mail: jiezhang2012@sjtu.edu.cn}}





\section{Introduction}\label{sec1}

Detailed balance (DB) is one of the most fundamental principles of science. 
In detailed-balanced systems, all kinetic processes are balanced on experimentally measurable time- and length-scales. For example, in mixtures undergoing chemical reactions at thermal equilibrium, each and every reaction $A+B\to C$, is balanced by the reverse reaction, $C\to A+B$, in the thermodynamic limit.
Since its early use by Maxwell~\cite{Max67} and formulation  by Boltzmann~\cite{Bo1872}, it has crystallised into two statements:
(i) In thermal equilibrium, each kinetic process is balanced, on average, by the reverse process, e.g., ${A}+{B}\leftrightharpoons{C}$. 
(ii) Detailed balance is unique to equilibrium systems and balance in steady states of non-equilibrium systems can only be achieved through cycles, e.g., ${A}\to{B}\to{C}\to{A}$~\cite{Klein}.
This principle was used extensively by later luminaries, including Einstein, Kramers, Eddington, Pauli, Ehrenfest, Dirac, Onsager and others, to model systems and kinetic processes that advanced greatly natural sciences~\cite{Ma1867,EinsteinLaser,Pa23,EiEh23,Di24,On31}. This principles is currently part of the established scientific paradigm.

We report here careful experimental observations on a family of non-equilibrium dynamics that cast doubts on the general validity of the above two statements. Specifically, we show that DB is satisfied in these non-equilibrium systems in spite of the fact that they could support, in principle, kinetic cycles. 
Another aim of these experiments is to gain insight into the structural self-organization of granular systems. The reason is that the structure, even on the few particles scale, plays an essential role in determining the large-scale mechanical and transport properties. 

A key structural characteristic in granular systems is the contact network, whose nodes are the intergranular contacts. In quasistatic processes, the structure evolves through contact making and breaking and the statistics of the contact network determine much of the bulk properties. 
Yet, the hysteretic and dissipative nature of the dynamics, combined with the discontinuous nature of the intergranular interactions, has hindered both predictive modelling and interpretations of experimental observations. Quantifying contact dynamics is significant to understanding granular dynamics and some attempts have been made in this direction~\cite{saitoh2015master,saitoh2019transition}. 

In planar granular systems, the contact network defines cells, which are the smallest closed loops of contacts around voids. Defining a cell's order as the number of particles enclosing it, the cell order distribution (COD) has proved to be key to understanding many properties of such systems~\cite{Toetal11,RiFo13,RafiTakashiGranMatt,BlRCP21}. In particular, it has been shown recently that, in steady states of quasistatic dynamics of dense granular systems, the COD maximises the entropy~\cite{Competition}. This observation is significant because the topology (or connectivity) has been shown to dominate the entropy in granular systems~\cite{AmBl17}. These findings extend the pioneering idea of Edwards and coworkers that granular materials can be described using the statistical-mechanics framework ~\cite{Edwards1,Edwards2,EdwardsReview}.
The COD has also been used recently to solve the old random close packing problem in two dimensions~\cite{BlRCP21}.

Breaking a contact between two (rattler-free) cells of order $i$ and $j$ merges them into a cell of order $\left(i+j-2\right)$. This kinetic process and its reverse are denoted $c_i + c_j \leftrightharpoons c_{i+j-2}$ and a sketch of it is shown in Fig.~\ref{fig1}A for $i=5$ and $j=6$. This process is analogous to  a chemical reaction and its reverse. The net flow of the process is described by a balance parameter~\cite{RafiTakashiGranMatt}
\begin{equation}
\eta_{i,j}\equiv p_{i,j}Q_i Q_j - q_{i+j-2,i}Q_{i+j-2} \equiv \eta_{M;i,j} - \eta_{B;i,j},
\label{etaij}
\end{equation}
with $Q_k$ the fraction of cells of order $k$.
When $\eta_{M;i,j}>\eta_{B;i,j}$ more cells are merging than breaking and vice versa. Rattlers, which are particles that do not transmit forces, can be neglected in the analysis of such systems, as has been discussed in~\cite{Competition}. The evolution equations of the COD,
\begin{equation}
\dot{Q}_k = f_k\left(\left\{\eta_{i,j}\right\}, \left\{Q_j\right\}\right) \ ,
\label{Evolution}
\end{equation}
and their steady states have been analysed in \cite{RafiTakashiGranMatt}. In particular, it was shown that steady states of very dense 2D systems (cell orders $<6$) have at most two processes and therefore can have no cycles, which means that, by default, they satisfy DB, $\eta_{i,j}=0$ for all $i,j\leq5$~\cite{RafiLim}.  

Here, we present experimental evidence that DB is satisfied much more generally in cyclically sheared granular systems of arbitrary density and upper cell order. By  carefully generating very loose assemblies with cell orders of over $30$ and testing the balance of the $\eta_{i,j}$-processes up to order $10$, we show that $\eta_{B;i,j}=\eta_{M;i,j}$ in each and every process.

\section{Results}\label{sec2}

\begin{figure}[htb]
	\centerline{\includegraphics[width=8.6cm]{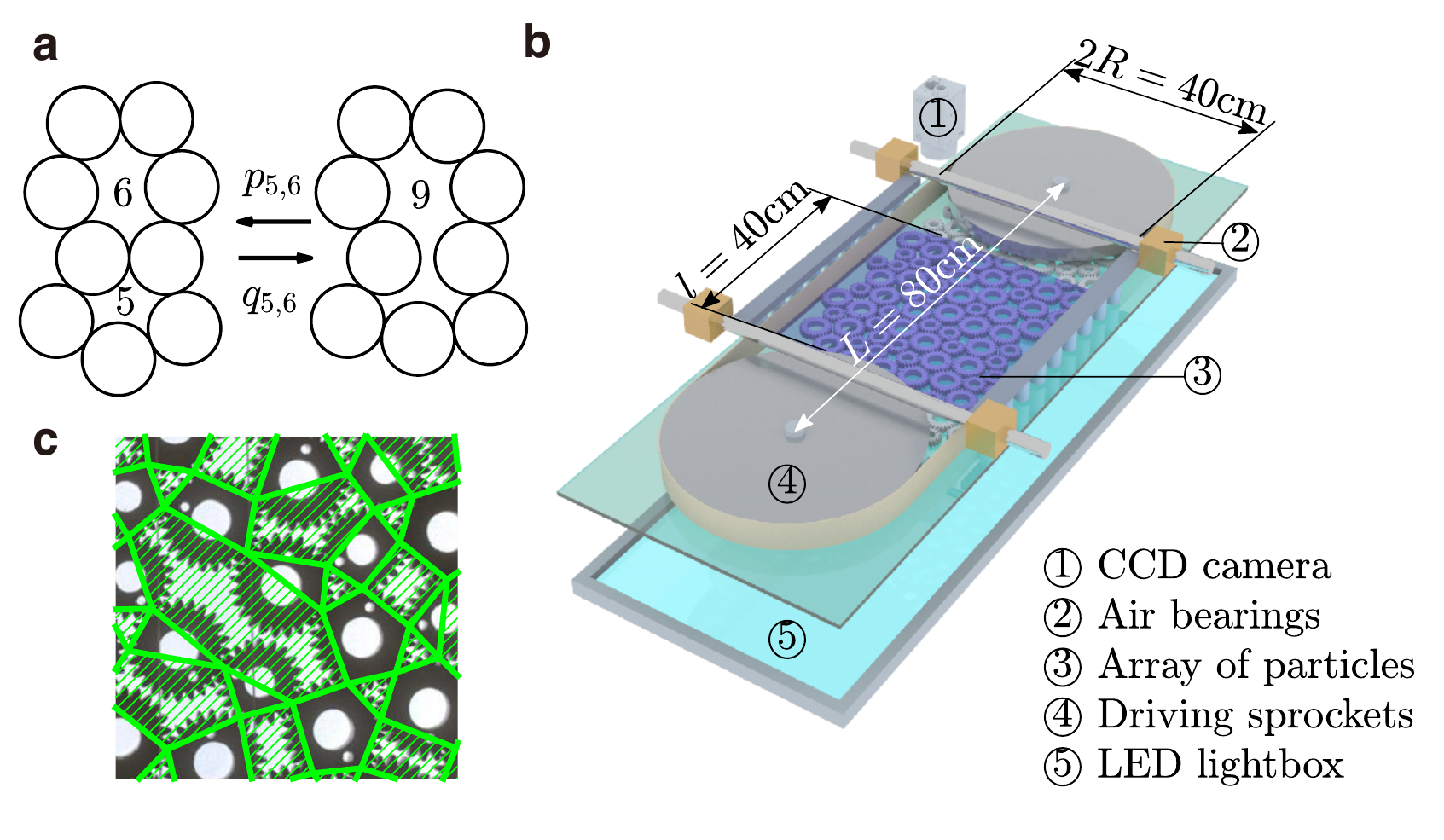}}
	\caption{\textbf{Sketch of cells and experimental setup}\label{fig1}
		(A) The processes of cells breaking and merging, expmplified for $c_5 + c_6 \leftrightharpoons c_{9}$.
		(B) Sketch of stadium shear device, which cyclic shears the granular system within under a constant boundary pressure.
		(C) A typical part of the gear systems -- cells are shaded green.}
\end{figure}

The experimental setup, known as the \emph{stadium shear device}~\cite{Itai,Competition}, is sketched in Fig.~\ref{fig1}B. This device mimics simple shear between two infinite parallel plates. We used a stepping motor to drive periodically two stainless steel sprockets, connected by a rubber belt, corrugated on the inside to ensure that particles do not slip against it. Monitoring the particles within the region shaded blue in Fig.~\ref{fig1}B, we applied quasi-static cyclic strains, ranging from $2.5\%$ to $10\%$, to several systems of particles with a wide range of friction coefficient $\mu$: nylon gears ($\mu\to\infty$)~\cite{Corey}, photoelastic disks ($\mu\approx0.7$)~\cite{JiePEFriction}, ABS plastic cylindrical rings ($\mu\approx0.32$), stainless steel disks ($\mu\approx0.3$), and Teflon-coated photoelastic disks ($\mu\approx0.15$)~\cite{Competition}. The use of high-friction particles allowed us to explore low-density systems in mechanical equilibrium, down to packing fractions of $74\%$ (excluding rattlers) and containing cells up to order $30$. We verified that in all the systems, the particle-base friction is negligibly smaller than the interparticle one~\cite{Competition}.
The range of shear strains was chosen to ensure maximum identification of contacts and fastest decay of structural correlations over fewer than $20$ shear cycles for all orders~\cite{RenJiePRL}.
To minimize crystallization, we used a 50\%-50\% binary mixture of particles of size ratio 1:1.4. The small diameter was $1.0$cm in all the systems, except for the gears, in which the pitch diameter of the small particles was $1.6$cm and the tooth height $0.36$cm. Each system had about $2000$ particles. 
The packing fractions ranged from extremely dilute in the gear system, $0.74\pm0.01$, to $0.83\pm0.01$ for Teflon-coated photoelastic, stainless steel, and plastic particles.
Each experiment started by depositing particles as randomly as possible inside the device and applying cyclic shears quasi-statically. It has been shown~\cite{Competition} that these systems are liquid-like and that the packing fractions reach steady states within about $100$ shear cycles. To make sure that we study steady states, we applied several hundreds of shear cycles before collecting data. 

At the maximum strain of each cycle, we paused the shearing and took a snapshot of the structure, which allowed us a stroboscopic study of the dynamics. 
Below, we display the results for a strain amplitude of $5$\%, which are typical for all the strains we used. 
For each particle configuration and constructed the cells, as sketched in Fig.~\ref{fig1}(b-c)~\cite{RafiTakashi2014,RafiTakashi2017}. 

To quantify the error in contact identification, we used the thresholding method, detailed in the supplementary materials of~\cite{Competition} and varied the threshold to make sure that our results and conclusions are insensitive to the threshold.

\begin{figure}
	\centerline{\includegraphics[width = 8.6 cm]{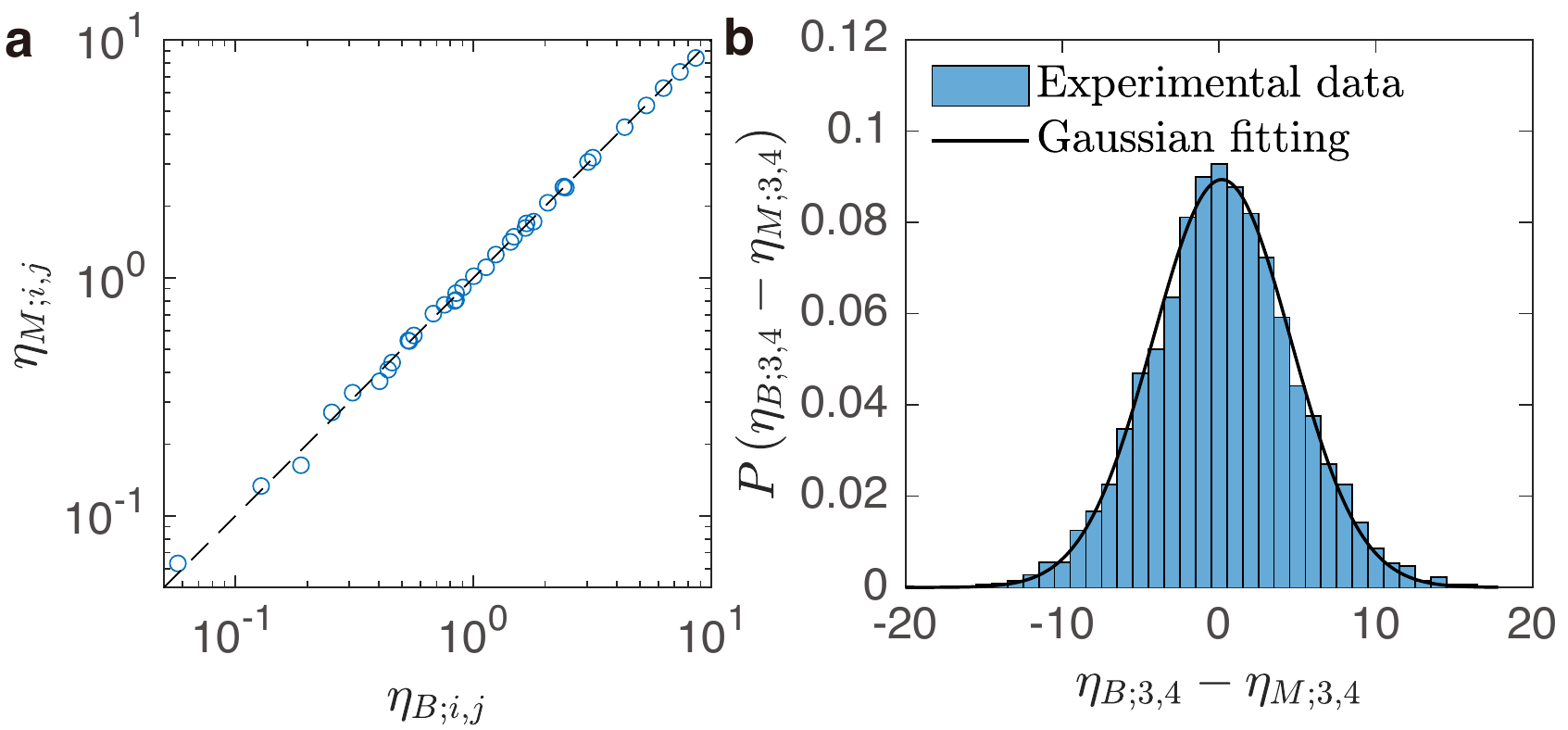}}
	\caption{\label{fig2}
		\textbf{Detailed balance for different cell orders} 
		(A) A plot of $\eta_{B;i,j}$ vs. $\eta_{M;i,j}$. Each point represents a specific $(i,j)$ pair. The plot shows an overall DB, $\eta_{B;i,j} = \eta_{M;i,j}$, to a very good accuracy.
		(B) A typical spread of the deviation from DB for  $i=3$, $j=4$, $\eta_{B;3,4} - \eta_{M;3,4}$, across 4000 shear cycles. The spread is well-fitted by a Gaussian of mean zero, establishing the convergence to DB. The convergence process is illustrated in Fig.\ref{fig3}}
\end{figure}

As in previous experiments~\cite{Competition}, we found that the CODs of the steady state in these systems are exponential. We calculated the average rates of $\eta_{B;i,j}$ and $\eta_{M;i,j}$ over all $(i,j)$ pairs, for each combination of $i$ and $j$, and averaged over thousands of shear cycles. The result for the gear systems, averaged over $4000$ cycles, is shown in Fig.\ref{fig2}A for all $i,j\leq10$, with each point representing one pair. The low number of cells of orders $>10$ prevents reliable statistics and the results for these orders are omitted from the plot. The plot shows clearly that these systems satisfy DB. 

As expected, the finite size of our systems must give rise to deviations from exact DB within any one cycle and an example of the distribution of the deviation $\eta_{B;i,j} - \eta_{M;i,j}$ across 4000 shear cycles is shown in Fig.~\ref{fig2}B for $(i,j)=(3,4)$, for which we have the best statistics. This distribution is fitted by a Gaussian of mean zero to a very good accuracy. Similar distributions are observed in the processes of all the other orders. 
The observation of the zero mean substantiates the conclusion that all these steady states satisfy DB.  To solidify this conclusion, we next investigated the convergence to DB as a function of the number of monitored cells and system size.

In all systems, the cell survival probability decays exponentially with the number of shear cycles, $e^{-N/\tau}$, with $\tau<2$ and even the longest lasting cells disappear after fewer than $20$ shear cycles~\cite{Competition}. States are practically uncorrelated when the survival probability of all cells falls below $10^{-3}$. The number of shear cycles over which this happens, $N_c$, depends weakly on the interparticle friction. This observation is important -- it allowed us to regard groups of $N_c$ shear cycles as independent systems, improving the  statistics of our data. Moreover, the absence of correlations betsween groups of shear cycles larger than $N_c$ means that a sub-ensemble of $N>N_c$ shear cycles is equivalent to studying $N/N_c$ independent systems. 
In turn, this is then equivalent to studying $N/N_c$ independent domains in a larger system whose linear size is at least $\sqrt{N/N_c}$ larger than our experimental one.

\begin{figure}[tb]
	\centerline{\includegraphics[width = 8.6 cm]{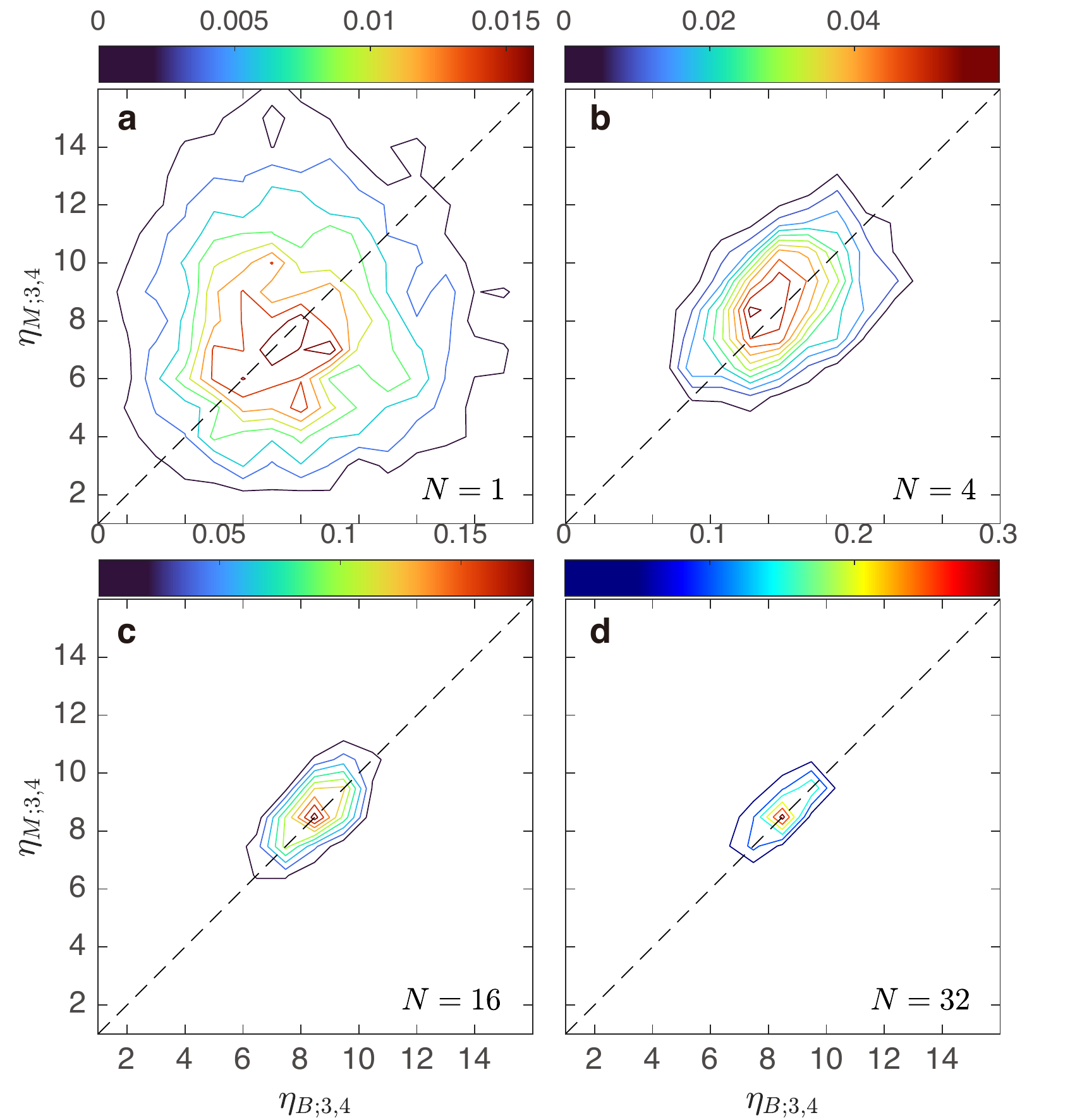}}
	\caption{\label{fig3}
	\textbf{Convergence to detailed balance}
Contour plots of the values of the pairs $\left(\eta_{B;3,4},\eta_{M;3,4}\right)$ for the number of shear cycles $N=1,4,16$, and $32$. The region spanned by the contours, whose values increase towards the DB line, shrinks rapidly as $N$ increases.}
\end{figure}
\begin{figure}[tb]
	\centerline{\includegraphics[width = 8.6 cm]{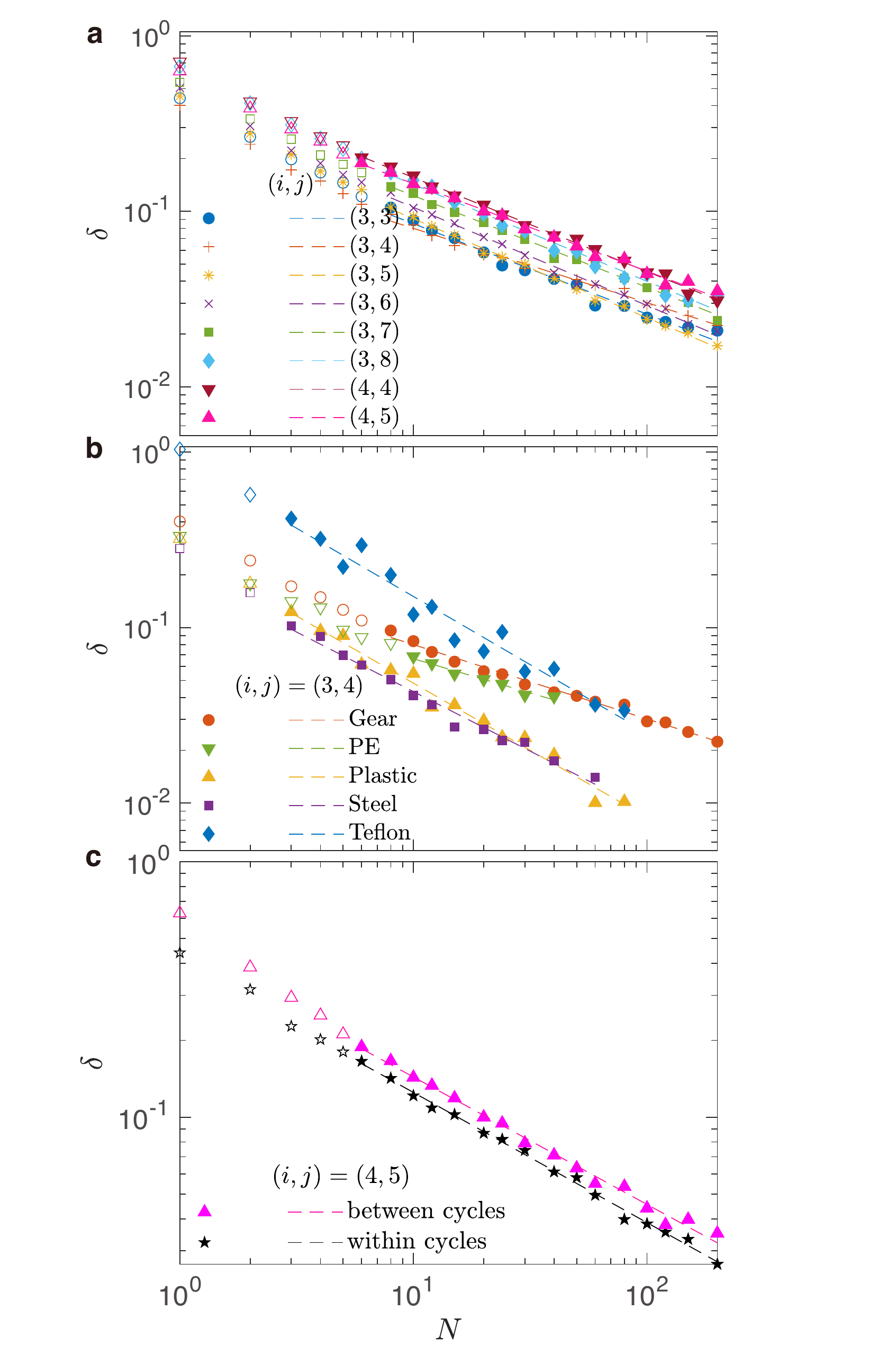}}
	\caption{\label{fig4}
		\textbf{The convergence to detailed balanced as a function of system size. }
		(A) The decrease of the deviation from perfect detailed balance in the gear systems as a function of the number of shear cycles $N$ for all the eight processes involving $3$- and $4$-cells. The decrease is algebraic, $\delta_{i,j}\sim N^{-\beta}$, with $\beta <1$, but dependent somewhat on the specific process. 
		(B) The typical convergence of a specific process, here $3+4\leftrightharpoons5$, in  systems, indexed by decreasing friction. The full symbols are for data collected from uncorrelated configurations, $N>N_c$. The decay is algebraic with $\beta$ increasing with decreasing interparticle friction. The dashed lines are fits to the uncorrelated sub-systems, $N>N_c$.
		(C) Comparison between data collected once a cycle (triangles) and using 16 frames within each cycle (stars). The difference in the value of $\beta$ between the two is only $5\%$ and confirms the convergence.}
\end{figure}

To investigate the rate of convergence to DB as a function of $N$, we divided the $4000$ shear cycles into $K=4000/N$ sub-ensembles, each of size $N=1,2,...,4000$. In each sub-ensemble, we calculated the averages of $\eta_{B;3,4}$ and $\eta_{M;3,4}$, obtaining $4000/N$ values of the pairs $\left(\eta_{B;3,4},\eta_{M;3,4}\right)$. A contour plot was then generated for these values, shown in Fig. \ref{fig3} for $N=1,4$, $16$, and $32$.
The contours, whose values increase towards the centre, are distributed symmetrically around the DB line $\eta_{B}=\eta_{M}$ and they narrow rapidly with $N$. The convergence to DB is very good already for $N=16$. We define the deviation from the DB line as
\begin{equation}
\delta_{i,j}=\frac{\sqrt{\langle\left(\overline{\eta_{B;i,j}}-\overline{\eta_{M;i,j}}\right)^2/2\rangle}}{\langle\overline{\eta_{B;i,j}}+\overline{\eta_{M;i,j}}\rangle/2} \ ,
\end{equation}
in which, $\overline{\eta}$ is the average within a sub-system of size $N$ and $\langle...\rangle$ is an ensemble average over all windows of size $N$. 
In Fig. \ref{fig4} we show the $N$-dependent rates of convergence of different subsets: Fig. \ref{fig4}A - the convergence rate of eight processes, up to $i,j=5$, in one system ($\mu\to\infty$) gear system; Fig. \ref{fig4}B - the approach of a typical process in systems of different friction coefficients. These firmly substantiate that the steady states satisfy an overall DB in the large-system limit. Interestingly, the convergence is always algebraic, $\delta_{i,j}\sim N^{-\beta}$, with the values of $\beta$ are less than unity. Overall, $d\beta/d\mu < 0$, i.e., the less frictional the particles the quicker the steady state is achieved. This is consistent with the observations that cells survival time gets shorter with decreasing $\mu$~\cite{Competition}.
To test the results for the effect of unobserved events within cycle, we took 16 snapshots within each cycle, identified all the events and re-calculated the deviation $\delta(N)$. The comparison is shown in Fig.\ref{fig4}C. The difference between the values of $\beta$ is $5$\%, with the values of $\delta$ slightly lower within the additional processes. This suggests that the convergence to DB is slightly faster when adding those processes. 

The observation of DB in such a wide range of systems suggests that DB in steady states of quasi-static cyclic shearing is robust and independent of contact-level characteristics and particle stiffness.

\section{Discussion and conclusion}\label{sec3}

To conclude, we have established that the steady state dynamics of the cell order distribution in cyclically sheared planar disc systems satisfy detailed balance. This result, which is valid for very low-density systems with cell orders at least as high as $10$, goes beyond validating the theoretical derivation in~\cite{RafiTakashiGranMatt}, which was shown to be limited to very dense systems that could not support kinetic cycle processes~\cite{RafiLim}. Our discovery of detailed balance in  non-equilibrium systems, which could very well support cycle processes, suggests that the existing paradigm about the exclusivity of this principle to equilibrium systems~\cite{Klein} must be reconsidered.

On a more fundamental level, DB has always been associated with existence of time-reversal symmetry~\cite{Ma1867,On31,EinsteinLaser}. Yet, in all our finite-friction systems, this symmetry is broken when the ratio of the tangential to normal forces exceeds a threshold and sliding ensues. The independence of DB of contact level details then suggests that it requires no time-reversal symmetry in these out-of-equilibrium systems, in stark contrast to DB in equilibrium systems.

In addition to establishing the robust occurrence of DB, we have investigated the approaches to these steady states and showed that it is algebraic in the number of cycles. We also argued that this is equivalent to convergence with system size and, therefore, that our conclusions can be safely extended to systems much larger than in our experiments. Nevertheless, the rapid convergence to these steady states, over less than $20$ cycles, implies that this is a robust result that can be observed at relatively small systems.

Finding detailed balance in out-of-equilibrium granular dynamics provides further supports Edwards's original conjecture that granular systems can be usefully studied with the statistical mechanics formalism~\cite{Edwards1,Edwards2,EdwardsReview,Rafi2003,Bletal16}. 
This should encourage further utilization of statistical mechanics to model the rich dynamic behavior of this form of matter. In particular, in view of the significant impact of the detailed balance principle to fundamental modelling equilibrium systems, we believe that it can similarly lead to many advances in non-equilibrium granular systems.

\section*{Author Contributions}
This work is supported by the NSFC (No.11974238 and No.11774221 ). This work is also supported by the Innovation Program of Shanghai Municipal Education Commission under No 2021-01-07-00-02-E00138.
J.Z. and R.B. conceived and supervised the project. X.L.S. and J.Z. designed the experiment. X.L.S. performed the experiment. Y.Q.W., Y.J.W., R.B. and J.Z. participated in the discussions. X.L.S., R.B. and J.Z. wrote the paper.

\section*{Conflicts of interest}
There are no conflicts to declare.

\section*{Acknowledgements}
X.L.S. and J.Z. acknowledge the support from the Student Innovation Center of Shanghai Jiao Tong University. X.L.S. and J.Z. also acknowledge the helpful discussion with Walter Kob. R.B. acknowledges the hospitality of the Cavendish Laboratory, University of Cambridge, UK. 



\balance


\bibliography{ref2}
\bibliographystyle{rsc}

\end{document}